\documentclass[a4paper,11pt]{article}
\usepackage{amsmath,amssymb,amsfonts,amsthm}
\usepackage{a4wide}

\newcommand{\com}[2]{[#1,#2]}
\newcommand{\starcom}[2]{[#1\stackrel{\star}{,}#2]}
\newtheorem{propo}{Proposition}

\title{%
  Symmetry~Reduction in Twisted~Noncommutative~Gravity
  with~Applications~to~Cosmology~and~Black~Holes}

\author{%
  Thorsten Ohl%
    \thanks{e-mail: \texttt{ohl@physik.uni-wuerzburg.de}}\\
  Alexander Schenkel%
    \thanks{e-mail: \texttt{aschenkel@physik.uni-wuerzburg.de}}\\
  \hfil\\
  Institut f\"ur Theoretische Physik und Astrophysik\\
  Universit\"at W\"urzburg, Am Hubland, 97074 W\"urzburg, Germany}

\date{October 2008}
\begin{document}
\maketitle
\begin{abstract}
  As a preparation for a mathematically consistent study of the
  physics of symmetric spacetimes in a noncommutative setting, we
  study symmetry reductions in deformed gravity.  We focus on
  deformations that are given by a twist of a Lie algebra acting on the
  spacetime manifold. We derive conditions on those twists that allow a
  given symmetry reduction. A complete classification of admissible
  deformations is possible in a class of twists generated by
  commuting vector fields.  As examples, we explicitly construct the families of
  vector fields that generate twists which are compatible with
  Friedmann-Robertson-Walker cosmologies and Schwarzschild black
  holes, respectively.  We find nontrivial isotropic twists of FRW
  cosmologies and nontrivial twists that are compatible with all
  classical symmetries of black hole solutions.
\end{abstract}
\renewcommand{\arraystretch}{1.5}

\section{Introduction}
The study of noncommutative geometry is an active topic in both
theoretical physics and mathematics. From the mathematical perspective
it is a generalization of classical (commutative) geometry.  From the
physics perspective it is suggested by the \textit{Gedankenexperiment}
of localizing events in spacetime with a Planck scale
resolution~\cite{Doplicher:1994zv}.  In this
\textit{Gedankenexperiment}, a sharp localization induces an
uncertainty in the spacetime coordinates, which can naturally be
described by a noncommutative spacetime. Furthermore, noncommutative
geometry and quantum gravity appear to be connected strongly and one
can probably model ``low energy'' effects of quantum gravity theories
using noncommutative geometry.

There have been many attempts to formulate scalar, gauge and gravity
theories on noncommutative spacetime, in particular using the simplest
example of a Moyal-Weyl spacetime having constant noncommutativity
between space and time coordinates, see
\cite{Szabo:2001kg,MullerHoissen:2007xy} for reviews.  Furthermore,
this framework had been applied to phenomenological particle physics
with~\cite{NCSM,NCSM-Pheno} and without Seiberg-Witten maps (see the
review~\cite{NC-Pheno} and references therein),
cosmology~\cite{NC-Cosmo} and black hole physics (see the
review~\cite{Nicolini:2008aj} and references therein).

Our work is based on the approach outlined
in~\cite{Aschieri:2005yw,Aschieri:2005zs,Aschieri:2006kc}, where a
noncommutative gravity theory based on an arbitrary twist deformation
is established. This approach has the advantages of being formulated
using the symmetry principle of deformed diffeomorphisms, being
coordinate independent and applicable to nontrivial topologies.
However, there is also the disadvantage that it does not match
the Seiberg-Witten limit of string theory~\cite{AlvarezGaume:2006bn}.
Nevertheless, string theory is not the only candidate for a
fundamental theory of quantum gravity. Therefore, the investigation of
deformed gravity remains interesting on its own terms and it could very well
emerge from a fundamental theory of quantum gravity different from
string theory.

The outline of this paper is as follows. In section \ref{sec:basics}
we review the basics of the formalism of twisted noncommutative
differential geometry. For more details and the proofs we refer to the
original paper~\cite{Aschieri:2005zs} and the
review~\cite{Aschieri:2006kc}. We will work with a general twist and
do not restrict ourselves to the Moyal-Weyl deformation.

In section \ref{sec:symred} we will study symmetry reduction in
theories based on twisted symmetries, such as the twisted
diffeomorphisms in our theory of interest. The reason is that we aim
to investigate which deformations of cosmological and black hole
symmetries are possible. We will derive the conditions that the twist
has to satisfy in order to be compatible with the reduced symmetry.
In section \ref{sec:jambor} we restrict the twists to the class of
Reshetikhin-Jambor-Sykora twists~\cite{Reshetikhin:1990ep,Jambor:2004kc}, 
that are twists generated by commuting vector fields and are convenient 
for practical applications. Within this restricted class of twists we 
can classify more explicitly the possible deformations of Lie algebra 
symmetries acting on a manifold~$\mathcal{M}$.

In section \ref{sec:cosmo} and \ref{sec:blackhole} we apply the
formalism to cosmological symmetries as well as the black hole. We
classify the possible Reshetikhin-Jambor-Sykora deformations of these models and
obtain physically interesting ones.  In section \ref{sec:conc} we
conclude and give an outlook to possible further investigations.  In
particular possible applications to phenomenological cosmology and
black hole physics will be discussed.

\section{\label{sec:basics}Basics of Twisted Differential Geometry and Gravity}
In order to establish notation, we will give a short summary of the framework of twisted differential geometry and gravity. 
More details can be found in~\cite{Aschieri:2005yw,Aschieri:2005zs,Aschieri:2006kc}.

There is a quite general procedure for constructing noncommutative spaces and their corresponding symmetries
by using a twist. For this we require the following ingredients~\cite{Aschieri:2006kc}:
\begin{enumerate}
 \item a Lie algebra $\mathfrak{g}$
 \item an action of the Lie algebra on the space we want to deform
 \item a twist element $\mathcal{F}$, constructed from the generators of the Lie algebra $\mathfrak{g}$
\end{enumerate}
By a twist element we denote an invertible element of $U\mathfrak{g}\otimes U\mathfrak{g}$, where 
$U\mathfrak{g}$ is the universal enveloping algebra of $\mathfrak{g}$. $\mathcal{F}$ has to fulfill some
conditions, which will be specified later. The basic idea in the following is to combine any bilinear map with the inverse twist
and therefore deform these maps. This leads to a mathematically consistent deformed theory covariant under
the deformed transformations. We will show this now for the deformation of diffeomorphisms.

For our purpose we are interested in the Lie algebra of vector fields $\Xi$ on a manifold $\mathcal{M}$.
The transformations induced by $\Xi$ can be seen as infinitesimal diffeomorphisms. A natural action of these
transformations on the algebra of tensor fields $\mathcal{T}:=\bigoplus\limits_{n,m} \bigotimes^{n} \Omega \otimes \bigotimes^m \Xi  $
is given by the Lie derivative $\mathcal{L}$. $\Omega$ denotes the space of one-forms.

In order to deform this Lie algebra, as well as its action on tensor fields and the tensor fields themselves, we first have 
to construct the enveloping algebra $U\Xi$. This is the associative 
tensor algebra generated by the elements of $\Xi$ and the unit $1$, modulo the left and right ideals generated by
the elements $\com{v}{w}-v w + w v$. This algebra can be seen as a Hopf algebra by using the following 
coproduct $\Delta$, antipode $S$ and counit $\epsilon$ defined on the generators $u\in \Xi$ and $1$ by:
\begin{flalign}
\begin{array}{ll}
 \Delta(u) = u\otimes 1 + 1\otimes u, & \Delta(1) = 1\otimes 1~,\\
 \epsilon(u) = 0, & \epsilon(1) = 1~,\\
 S(u) = -u, & S(1) = 1~.
\end{array}
\end{flalign}
These definitions can be consistently carried over to the whole enveloping algebra demanding $\Delta$ and $\epsilon$
to be algebra homomorphisms and $S$ to be an anti-homomorphism, i.e.~for any two elements $\eta,\xi\in U\Xi$, we require
\begin{subequations}
\begin{flalign}
 &\Delta(\eta\xi) = \Delta(\eta)\Delta(\xi)~,\\
 &\epsilon(\eta\xi) = \epsilon(\eta) \epsilon(\xi)~,\\
 &S(\eta\xi)= S(\xi) S(\eta)~.
\end{flalign}
\end{subequations}
The action of the enveloping algebra on the tensor fields can be defined by extending the Lie derivative
\begin{flalign}
 \mathcal{L}_{\eta\xi} (\tau) := \mathcal{L}_{\eta}(\mathcal{L}_{\xi} (\tau))~,~\forall \eta,\xi\in U\Xi~,~\tau\in \mathcal{T}~.
\end{flalign}
This action is consistent with the Lie algebra properties, since $\mathcal{L}_{\com{u}{v}}(\tau) = \mathcal{L}_{uv}(\tau)-\mathcal{L}_{vu}(\tau)$
 for all $u,v\in\Xi$ by the properties of the Lie derivative.

The extension of the Lie algebra $\Xi$ to the Hopf algebra $(U\Xi,\cdot,\Delta,S,\epsilon)$, where $\cdot$ is the multiplication in $U\Xi$, 
can now be used in order to construct deformations of it. For the deformations we restrict ourselves
to twist deformations, which is a wide class of possible deformations. The reason is that for twist deformations the
construction of deformed differential geometry and gravity can be performed explicitly by only using properties of the twist, 
see~\cite{Aschieri:2005zs}. Other deformations require further investigations.

In order to perform the deformation we require a twist element $\mathcal{F}= f^{\alpha}\otimes f_{\alpha}\in U\Xi\otimes U\Xi$ 
(the sum over $\alpha$ is understood) fulfilling the following conditions
\begin{subequations}
\begin{flalign}
 &\mathcal{F}_{12}(\Delta\otimes \mathrm{id})\mathcal{F}=\mathcal{F}_{23}(\mathrm{id}\otimes\Delta)\mathcal{F}~,\\
 &(\epsilon\otimes \mathrm{id}) \mathcal{F} = 1 = (\mathrm{id}\otimes\epsilon)\mathcal{F}~,\\
 &\mathcal{F}=1\otimes1 +\mathcal{O}(\lambda)~,
\end{flalign}
\end{subequations}
where $\mathcal{F}_{12}:= \mathcal{F}\otimes 1 $, $\mathcal{F}_{23}:=1\otimes\mathcal{F}$ and $\lambda$ is the deformation parameter. 
The first condition will assure the associativity of the deformed products, the second will assure that deformed 
multiplications with unit elements will be trivial and the third condition assures the existence of the undeformed 
classical limit $\lambda\to 0$. 
Furthermore, we can assume without loss of generality that $f_\alpha$ (and also $f^\alpha$) are linearly independent for all $\alpha$, 
what can be assured by combining linearly dependent $f$.
Note that $\mathcal{F}$ is regarded as formal power series in $\lambda$, such as the deformation itself. 
Strict (convergent) deformations will not be regarded here.

The simplest example is the twist on $\mathbb{R}^n$ given by 
$\mathcal{F}_\theta:= \exp{\bigl(-\frac{i\lambda}{2}\theta^{\mu\nu} \partial_\mu\otimes\partial_\nu \bigr)}$ with $\theta^{\mu\nu}=\mathrm{const.}$
 and antisymmetric, leading to the Moyal-Weyl deformation, but there are also more complicated ones.

From a twist, one can construct the twisted triangular Hopf algebra
 $(U\Xi_\mathcal{F},\cdot,\Delta_\mathcal{F},S_\mathcal{F},\epsilon_\mathcal{F})$
with $R$-matrix $R:=\mathcal{F}_{21}\mathcal{F}^{-1}=:R^\alpha \otimes R_\alpha$, inverse $R^{-1} =: \bar R^\alpha \otimes \bar R_\alpha = R_{21}$ 
 and
\begin{flalign}
 &\Delta_\mathcal{F}(\xi):= \mathcal{F}\Delta(\xi)\mathcal{F}^{-1}~,\quad\epsilon_\mathcal{F}(\xi):= \epsilon(\xi)~,\quad S_\mathcal{F}(\xi):= \chi S(\xi)\chi^{-1}~,
\end{flalign}
where $\chi := f^\alpha S(f_\alpha)$, $\chi^{-1}:= S(\bar f^\alpha) \bar f_\alpha$ and $\bar f^\alpha \otimes \bar f_\alpha := \mathcal{F}^{-1}$.
Furthermore, $\mathcal{F}_{21}:= f_\alpha\otimes f^\alpha$ and $R_{21}:=R_\alpha \otimes R^\alpha$. Again, we can assume without loss of generality that all summands of $\mathcal{F}^{-1}$, $R$ and $R^{-1}$ are linearly independent.

However, as explained in~\cite{Aschieri:2005zs}, it is simpler to use the triangular $\star$-Hopf algebra $\mathcal{H}_\Xi^\star=(U\Xi_\star,\star,\Delta_\star,S_\star,\epsilon_\star)$,
isomorphic to $(U\Xi_\mathcal{F},\cdot,\Delta_\mathcal{F},S_\mathcal{F},\epsilon_\mathcal{F})$. The operations
in this algebra on its generators $u,v\in\Xi$ (note that this algebra has the same generators as the classical Hopf algebra) are defined by
\begin{subequations}
\label{eqn:defstarhopfactions}
\begin{flalign}
 &u\star v := \bar f^\alpha(u)\bar f_\alpha(v)~,\\ 
 &\Delta_\star(u) := u\otimes 1 + X_{\bar R^\alpha} \otimes \bar R_\alpha(u)~,\\
 &\epsilon_\star(u):=\epsilon(u)=0~,\\
 &S_\star^{-1}(u) := - \bar R^\alpha(u) \star X_{\bar R_\alpha}~,
\end{flalign}
\end{subequations}
where for all $\xi\in U\Xi$ we define $X_\xi := \bar f^\alpha \xi \chi S^{-1}(\bar f_\alpha)$. The action of the twist on the elements of $U\Xi$
is defined by extending the Lie derivative to the adjoint action~\cite{Aschieri:2005zs}. Note that $U\Xi=U\Xi_\star$ as vector spaces. The $R$-matrix is
 given by $R_\star := X_{R^\alpha} \otimes X_{R_\alpha}$ and is triangular. The coproduct and antipode (\ref{eqn:defstarhopfactions}) 
is defined consistently on $U\Xi_\star$ by using for all $\xi,\eta\in U\Xi_\star$ the definitions
\begin{flalign}
 \Delta_\star(\xi\star\eta):= \Delta_\star(\xi)\star\Delta_\star(\eta)~,\quad~S_\star(\xi\star\eta):= S_\star(\eta)\star S_\star(\xi)~.
\end{flalign}

The next step is to define the $\star$-Lie algebra of deformed infinitesimal diffeomorphisms. It has been shown~\cite{Aschieri:2005zs} 
that for the twist deformation case the choice $(\Xi_\star,\com{~}{~}_\star)$, where $\Xi_\star=\Xi$ as vector spaces and 
\begin{flalign}
\com{u}{v}_\star:= \com{\bar f^\alpha(u)}{\bar f_\alpha(v)} 
\end{flalign}
is a natural choice for a $\star$-Lie algebra. It fulfills all conditions which are necessary for a sensible $\star$-Lie algebra given by
\begin{enumerate}
 \item $\Xi_\star\subset U\Xi_\star$ is a linear space, which generates $U\Xi_\star$
 \item $\Delta_\star(\Xi_\star) \subseteq \Xi_\star\otimes 1 + U\Xi_\star\otimes\Xi_\star$
 \item $\com{\Xi_\star}{\Xi_\star}_\star\subseteq\Xi_\star$
\end{enumerate}
The advantage of using the $\star$-Hopf algebra $(U\Xi_\star,\star,\Delta_\star,S_\star,\epsilon_\star)$ 
instead of the $\mathcal{F}$-Hopf algebra $(U\Xi_\mathcal{F},\cdot,\Delta_\mathcal{F},S_\mathcal{F},\epsilon_\mathcal{F})$
is that the $\star$-Lie algebra of vector fields is isomorphic to $\Xi$ as a vector space.
For the $\mathcal{F}$-Hopf algebra this is not the case and the $\mathcal{F}$-Lie algebra consists in general of multidifferential
 operators.

The algebra of tensor fields $\mathcal{T}$ is deformed by using the $\star$-tensor product~\cite{Aschieri:2005zs}
\begin{flalign}
 \tau \otimes_\star \tau^\prime := \bar f^\alpha(\tau) \otimes\bar f_\alpha(\tau^\prime)~,
\end{flalign}
where as basic ingredients the deformed algebra of functions $A_\star:=(C^\infty(M),\star)$ as well as the $A_\star$-bimodules of
vector fields $\Xi_\star$ and one-forms $\Omega_\star$ enter. We call $\mathcal{T}_\star$ the deformed algebra of tensor fields. Note that
 $\mathcal{T}_\star = \mathcal{T}$ as vector spaces.

The action of the deformed infinitesimal diffeomorphisms on $\mathcal{T}_\star$ is defined by the $\star$-Lie derivative
\begin{flalign}
\label{eqn:starliederivative}
 \mathcal{L}^\star_{u}(\tau):= \mathcal{L}_{\bar f^\alpha(u)}(\bar f_\alpha(\tau))~,~\forall \tau\in\mathcal{T}_\star~,~u\in\Xi_\star~,
\end{flalign}
which can be extended to all of $U\Xi_\star$ by $\mathcal{L}^\star_{\xi\star\eta}(\tau):= \mathcal{L}^\star_\xi(\mathcal{L}^\star_\eta (\tau)) $.

Furthermore, we define the $\star$-pairing $\langle \cdot,\cdot\rangle_\star:\Xi_\star\otimes_\mathbb{C} \Omega_\star\to A_\star$ between 
vector fields and one-forms as
\begin{flalign}
\label{eqn:starpairing}
 \langle v,\omega\rangle_\star := \langle \bar f^\alpha(v),\bar f_\alpha(\omega)\rangle~,~\forall v\in\Xi_\star,~\omega\in\Omega_\star~,
\end{flalign}
where $\langle \cdot,\cdot\rangle$ is the undeformed pairing.

Based on the deformed symmetry principle one can define covariant derivatives, torsion and curvature. This leads to deformed Einstein
equations, see~\cite{Aschieri:2005zs}, which we do not have to review here, since we do not use them in the following.

\section{\label{sec:symred}Symmetry Reduction in Twisted Differential Geometry}
Assume that we have constructed a deformed gravity theory based on a twist $\mathcal{F}\in U\Xi\otimes U\Xi$. Like in Einstein
gravity, the physical applications of this theory is strongly dependent on symmetry reduction. In this section we first define 
what we mean by symmetry reduction of a theory covariant under a Lie algebraic symmetry (e.g. infinitesimal diffeomorphisms) 
and then extend the principles to deformed symmetries and $\star$-Lie algebras.

In undeformed general relativity we often face the fact that the systems we want to describe have certain (approximate) symmetries.
Here we restrict ourselves to Lie group symmetries.
For example in cosmology one usually constrains oneself to fields invariant under certain symmetry groups $G$, like e.g.~the euclidian
group $E_3$ for flat universes or the $SO(4)$ group for universes with topology $\mathbb{R}\times S_3$, where the spatial hypersurfaces 
are 3-spheres.
For a non rotating black hole one usually demands the metric to be stationary and spherically symmetric. Practically, one
uses the corresponding Lie algebra $\mathfrak{g}$ of the symmetry group $G$, represents it faithfully on the Lie algebra 
of vector fields $\Xi$ on the manifold $\mathcal{M}$ and demands the fields $\tau\in\mathcal{T}$, which occur in the theory, 
to be invariant under these transformations, i.e. we demand
\begin{flalign}
 \mathcal{L}_v(\tau)=0~,~\forall v\in \mathfrak{g}~.
\end{flalign}
Since the Lie algebra $\mathfrak{g}$ is a linear space we can choose a basis $\lbrace t_i :i=1,\cdots,\mathrm{dim}(\mathfrak{g})\rbrace$
 and can equivalently demand
\begin{flalign}
 \mathcal{L}_{t_i}(\tau)=0~,~\forall i=1,2,\cdots ,\mathrm{dim}(\mathfrak{g})~.
\end{flalign}
The Lie bracket of the generators has to fulfill
\begin{flalign}
 \com{t_i}{t_j}=f_{ij}^{~~k}t_k~,
\end{flalign}
where $f_{ij}^{~~k}$ are the structure constants.

One can easily show that if we combine two invariant tensors with the tensor product, the resulting tensor is invariant too
because of the trivial coproduct
\begin{flalign}
 \mathcal{L}_{t_i}(\tau\otimes\tau^\prime) = \mathcal{L}_{t_i}(\tau) \otimes \tau^\prime + \tau\otimes\mathcal{L}_{t_i}(\tau^\prime)~.
\end{flalign}
The same holds true for pairings $\langle v,\omega\rangle$ of invariant objects $v\in\Xi$ and $\omega\in\Omega$.

Furthermore, if a tensor is invariant under infinitesimal transformations, it is also invariant under (at least small) finite transformations, 
since they are given by exponentiating the generators. The exponentiated generators are part of the enveloping algebra, 
i.e.~$\exp(\alpha^i t_i)\in U\mathfrak{g}$, where $\alpha^i$ are parameters. For large finite transformations the topology of the 
Lie group can play a role, such that the group elements may not simply be given by exponentiating the generators. In the following 
we will focus only on small finite transformations in order to avoid topological effects.

We now generalize this to the case of $\star$-Hopf algebras and their corresponding $\star$-Lie algebras. Our plan is
as follows: we start with a suitable definition of a $\star$-Lie subalgebra constructed from the Lie algebra $(\mathfrak{g},\com{~}{~})$.
This definition is guided by conditions, which allow for deformed symmetry reduction using infinitesimal transformations. Then
 we complete this $\star$-Lie subalgebra in several steps to a $\star$-enveloping subalgebra, a $\star$-Hopf subalgebra 
and a triangular $\star$-Hopf subalgebra. We will always be careful that the dimension of the $\star$-Lie subalgebra remains the same 
as the dimension of the corresponding classical Lie algebra. At each step we obtain several restrictions between the twist 
and $(\mathfrak{g},\com{~}{~})$.

We start by taking the generators $\lbrace t_i\rbrace$ of $\mathfrak{g}\subseteq\Xi$ 
and representing their deformations in the $\star$-Lie algebra $(\Xi_\star,\com{~}{~}_\star)$ as
\begin{flalign}
\label{eqn:stargenerators}
 t_i^\star = t_i +\sum\limits_{n=1}^{\infty} \lambda^n t_i^{(n)}~,
\end{flalign}
where $\lambda$ is the deformation parameter and $t_i^{(n)}\in\Xi_\star$.

The span of these deformed generators, together with the $\star$-Lie bracket, should form a $\star$-Lie subalgebra $(\mathfrak{g}_\star,\com{~}{~}_\star) := (\mathrm{span}(t_i^\star),\com{~}{~}_\star )$. Therefore $(\mathfrak{g}_\star,\com{~}{~}_\star)$ has
to obey certain conditions. Natural conditions are
\begin{subequations}
\label{eqn:infinitesimalconditions}
\begin{align}
 \label{eqn:infconda}\com{\mathfrak{g_\star}}{\mathfrak{g}_\star}_\star \subseteq \mathfrak{g}_\star,
   \quad &\text{i.e. $\com{t_i^\star}{t_j^\star}_\star =
     f_{ij}^{\star~k}t^\star_k$ with $f_{ij}^{\star~k} = f_{ij}^{~~k}
     + \mathcal{O}(\lambda)$} \\ \label{eqn:infcondb}
  \Delta_\star(\mathfrak{g}_\star) \subseteq
      \mathfrak{g}_\star \otimes 1 + U\Xi_\star \otimes
      \mathfrak{g}_\star,
   \quad &\text{which is equivalent to $\bar
     R_\alpha(\mathfrak{g}_\star)\subseteq
     \mathfrak{g}_\star~\forall_\alpha$}
\end{align}
\end{subequations}
The first condition is a basic feature of a $\star$-Lie algebra.
The second condition implies that if we have two $\mathfrak{g}_\star$ invariant tensors $\tau,\tau^\prime\in\mathcal{T}_\star$, the 
$\star$-tensor product of them is invariant as well
\begin{flalign}
 \mathcal{L}^\star_{t_i^\star}(\tau\otimes_\star\tau^\prime) = \mathcal{L}^\star_{t_i^\star}(\tau)\otimes_\star\tau^\prime + \bar R^\alpha(\tau)\otimes_\star\mathcal{L}^\star_{\bar R_\alpha (t_i^\star)}(\tau^\prime) =0~,
\end{flalign}
since $\bar R_\alpha(t_i^\star) \in \mathfrak{g}_\star$. 
The $\star$-pairings $\langle v,\omega\rangle_\star$ of two invariant objects $v\in\Xi_\star$ and $\omega\in\Omega_\star$ are also invariant
 under the $\star$-action of $\mathfrak{g}_\star$.
These are important features if one wants to combine invariant objects to e.g.~an invariant action.
Furthermore, the conditions are sufficient such that the following consistency relation is fulfilled for 
any invariant tensor $\tau\in\mathcal{T}_\star$
\begin{flalign}
 0=f_{ij}^{\star~k} \mathcal{L}^\star_{t^\star_k}(\tau) = \mathcal{L}^\star_{\com{t^\star_i}{t^\star_j}}(\tau)= \mathcal{L}^\star_{t^\star_{i}}(\mathcal{L}^\star_{t^\star_j}(\tau)) -   \mathcal{L}^\star_{\bar R^\alpha(t^\star_{j})}(\mathcal{L}^\star_{\bar R_\alpha(t^\star_i)}(\tau))~,
\end{flalign}
since $\bar R_\alpha(t_i^\star) \in \mathfrak{g}_\star$. 

Hence by demanding the two conditions (\ref{eqn:infinitesimalconditions}) for the $\star$-Lie subalgebra 
$(\mathrm{span}(t_i^\star),\com{~}{~}_\star )$ we can consistently perform symmetry reduction by using deformed 
{\it infinitesimal} transformations. In the classical limit $\lambda\to 0$ we obtain the classical Lie algebra 
$(\mathfrak{g}_\star,\com{~}{~}_\star ) \stackrel{\lambda\to 0}{\longrightarrow} (\mathfrak{g},\com{~}{~})$.

Next, we consider the extension of the $\star$-Lie subalgebra $(\mathfrak{g}_\star,\com{~}{~}_\star)\subseteq(\Xi_\star,\com{~}{~}_\star)$ 
to the triangular $\star$-Hopf subalgebra $\mathcal{H}_\mathfrak{g}^\star=(U\mathfrak{g}_\star,\star,\Delta_\star,S_\star,\epsilon_\star)\subseteq\mathcal{H}_\Xi^\star$.
This can be seen as extending the infinitesimal transformations to a quantum group. We will divide this path into several steps, 
where in every step we have to demand additional restrictions on the twist.

Firstly, we construct the $\star$-tensor algebra generated by the elements of $\mathfrak{g}_\star$ and $1$. We take this tensor algebra
modulo the left and right ideals generated by the elements $\com{u}{v}_\star - u\star v + \bar R^\alpha(v)\star \bar R_\alpha(u)$.
It is necessary that these elements are part of $U\mathfrak{g}_\star$, i.e.~we require
\begin{flalign}
\label{eqn:envelopcond}
 \bar R^\alpha(\mathfrak{g}_\star)\star \bar R_\alpha(\mathfrak{g}_\star)\subseteq U\mathfrak{g}_\star~.
\end{flalign}
This leads to the algebra $(U\mathfrak{g}_\star,\star)$, which is a subalgebra of $(U\Xi_\star,\star)$.

Secondly, we extend this subalgebra to a $\star$-Hopf subalgebra. Therefore we additionally have to require that 
\begin{subequations}
 \label{eqn:hopfconditions}
\begin{align}
 \label{eqn:hopfconditions1}\Delta_\star(U\mathfrak{g}_\star) &\subseteq U\mathfrak{g}_\star \otimes U\mathfrak{g}_\star~,\\
 \label{eqn:hopfconditions2}S_\star(U\mathfrak{g}_\star)&\subseteq U\mathfrak{g}_\star~.
\end{align}
\end{subequations}
Note that we do not demand that $S^{-1}_\star$ (defined on $U\Xi_\star$)
closes in $U\mathfrak{g}_\star$, since this is in general not the case for a nonquasitriangular Hopf algebra and we do not want to
demand quasitriangularity at this stage.
Then the $\star$-Hopf algebra $\mathcal{H}_\mathfrak{g}^\star$ is a Hopf subalgebra of $\mathcal{H}_\Xi^\star$.

Thirdly, we additionally demand that there exists an $R$-matrix $R_\star\in U\mathfrak{g}_\star\otimes U\mathfrak{g}_\star$. It is natural
to take the $R$-matrix of the triangular $\star$-Hopf algebra $\mathcal{H}_\Xi^\star$ defined by $R_\star := X_{R^\alpha}\otimes X_{R_\alpha}$. 
This leads to the restrictions
\begin{flalign}
\label{eqn:triangularcond}
X_{R^\alpha},X_{R_\alpha}\in U\mathfrak{g}_\star~,~\forall_\alpha.
\end{flalign}
Since $R_\star$ is triangular, i.e.~$R_\star^{-1 }=\bar R_\star^\alpha \otimes \bar R_{\star\alpha} = R_{\star21} = R_{\star\alpha} \otimes R_\star^\alpha$, we also have
$X_{\bar R^\alpha},X_{\bar R_\alpha}\in U\mathfrak{g}_\star~,~\forall_\alpha$.
If these conditions are fulfilled, $\mathcal{H}_\mathfrak{g}^\star$ is a triangular $\star$-Hopf subalgebra of $\mathcal{H}_\Xi^\star$ 
with the same $R$-matrix.

As we have seen, extending the $\star$-Lie subalgebra to a (triangular) $\star$-Hopf subalgebra gives severe restrictions on 
the possible deformations, more than just working with the deformed infinitesimal transformations given by a $\star$-Lie subalgebra 
or the finite transformations given by the $\star$-enveloping subalgebra $(U\mathfrak{g}_\star,\star)$.
Now the question arises if we actually require the deformed finite transformations to form a (triangular) $\star$-Hopf algebra in order to
use them for a sensible symmetry reduction. Because $(U\mathfrak{g}_\star,\star)$ describes deformed finite transformations and we
 have the relation
\begin{flalign}
\label{eqn:inf-fin}
\mathcal{L}^\star_{U\mathfrak{g}_\star \backslash \lbrace 1\rbrace}(\tau)=\lbrace0\rbrace\Leftrightarrow\mathcal{L}^\star_{\mathfrak{g}_\star}(\tau)=\lbrace0\rbrace~,
\end{flalign}
we can consistently demand tensors to be invariant under $(U\mathfrak{g}_\star,\star)$, since we
require tensors to be invariant under $(\mathfrak{g}_\star,\com{~}{~}_\star)$.
Therefore, a well defined $(U\mathfrak{g}_\star,\star)$ leads to a structure sufficient for symmetry reduction.
 The equivalence (\ref{eqn:inf-fin}) can be shown by using linearity
 of the $\star$-Lie derivative and the property $\mathcal{L}^\star_{\xi\star\eta}(\tau)=\mathcal{L}^\star_{\xi}(\mathcal{L}^\star_\eta(\tau))$.

In order to better understand the different restrictions necessary for constructing the $\star$-Lie subalgebra $(\mathfrak{g}_\star,\com{~}{~}_\star)$,
 the $\star$-enveloping subalgebra and the (triangular) $\star$-Hopf subalgebra $(U\mathfrak{g}_\star,\star,\Delta_\star,S_\star,\epsilon_\star)$, 
we restrict ourselves in the following sections to the class of Reshetikhin-Jambor-Sykora twists~\cite{Reshetikhin:1990ep,Jambor:2004kc}. This is a 
suitable nontrivial generalization of the Moyal-Weyl product, also containing e.g.~$\kappa$ and $q$ deformations when applied to Poincar\'{e} symmetry.

\section{\label{sec:jambor}The Case of Reshetikhin-Jambor-Sykora Twists}
Let $\lbrace V_a\in\Xi\rbrace$ be an arbitrary set of mutually commuting vector fields, 
i.e. $\com{V_a}{V_b}=0~,~\forall_{a,b}$, on an $n$ dimensional manifold $\mathcal{M}$. Then the object
\begin{flalign}
\label{eqn:jstwist}
 \mathcal{F}_{V} := \exp\bigl(-\frac{i\lambda}{2} \theta^{ab} V_a\otimes V_b \bigr)\in U\Xi\otimes U\Xi
\end{flalign}
is a twist element, if $\theta$ is constant and antisymmetric~\cite{Aschieri:2005zs,Reshetikhin:1990ep,Jambor:2004kc}.
 We call (\ref{eqn:jstwist}) a Reshetikhin-Jambor-Sykora twist. Note that this twist is not restricted to the topology $\mathbb{R}^n$ for the manifold
 $\mathcal{M}$.

Furthermore, we can restrict ourselves to $\theta$ with maximal rank and an even number of vector fields $V_a$, 
since we can lower the rank of the Poisson structure afterwards by choosing some of the $V_a$ to be zero.
We can therefore without loss of generality use the standard form
\begin{flalign}
\label{eqn:theta}
 \theta=\begin{pmatrix}
         0 & 1 & 0 & 0 & \cdots\\
	-1 & 0 & 0 & 0 & \cdots\\
	 0 & 0 & 0 & 1 & \cdots\\
	 0 & 0 & -1 & 0 & \cdots\\
	\vdots&\vdots&\vdots&\vdots&\ddots
        \end{pmatrix}
\end{flalign}
by applying a suitable $GL(n)$ transformation on the $V_a$.

This twist element is easy to apply and in particular we have for the inverse and the $R$-matrix
\begin{flalign}
\label{eqn:jsrmatrix}
 \mathcal{F}_V^{-1} = \exp\bigl(\frac{i\lambda}{2} \theta^{ab} V_a\otimes V_b \bigr)~,\quad~R=\mathcal{F}_{V,21}\mathcal{F}_V^{-1}=\mathcal{F}_V^{-2}=\exp\bigl(i \lambda\theta^{ab} V_a\otimes V_b \bigr)~.
\end{flalign}

Now let $(\mathfrak{g},\com{~}{~})\subseteq(\Xi,\com{~}{~})$ be the Lie algebra of the symmetry we want to deform. We choose a 
basis of this Lie algebra $\lbrace t_i : i=1,\cdots ,\mathrm{dim}(\mathfrak{g})\rbrace$ with $\com{t_i}{t_j}=f_{ij}^{~~k}t_k$.

Next, we discuss the symmetry reduction based on the $\star$-Lie subalgebra, as explained in section \ref{sec:symred}. 
Therefore we make the ansatz (\ref{eqn:stargenerators}) for the generators $t_i^\star$.
Furthermore, we evaluate the two conditions (\ref{eqn:infinitesimalconditions}) the $t_i^\star$ have to satisfy. 
We start with the coproduct condition (\ref{eqn:infcondb}), which
is equivalent to $\bar R_\alpha(t_i^\star)\in \mathrm{span}(t_i^\star),~ \forall_\alpha$, where $\alpha$ is a multi index. Using the
explicit form of the inverse $R$-matrix (\ref{eqn:jsrmatrix}) we arrive at the conditions 
\begin{flalign}
\label{eqn:condition1}
 \com{V_{a_1}}{\cdots\com{V_{a_n}}{t_i^\star}\cdots} = \mathcal{N}_{a_1\cdots a_n i}^{\star j} t_j^\star~,
\end{flalign}
where $\mathcal{N}_{a_1\cdots a_n i}^{\star j} :=\mathcal{N}_{a_1\cdots a_n i}^j + \sum\limits_{k=1}^{\infty} \lambda^k \mathcal{N}_{a_1\cdots a_n i}^{(k)~j}$ are constants. 

The only independent condition in (\ref{eqn:condition1}) is given by
\begin{flalign}
 \label{eqn:condition2}
\com{V_a}{t_i^\star} = \mathcal{N}^{\star j}_{a i} t^\star_j ~,
\end{flalign}
since it implies all the other ones by linearity. 
In particular, the zeroth order in $\lambda$ of (\ref{eqn:condition2}) yields
\begin{flalign}
\label{eqn:condition3}
 \com{V_a}{t_i} = \mathcal{N}_{ai}^{j} t_j~.
\end{flalign}

This leads to the following
\begin{propo}
\label{propo:ideal}
Let $(\mathfrak{g},\com{~}{~})\subseteq(\Xi,\com{~}{~})$ be a classical Lie algebra and $(\Xi_\star,\com{~}{~}_\star)$ the
 $\star$-Lie algebra of vector fields deformed by a Reshetikhin-Jambor-Sykora twist, constructed with vector fields $V_a$.
Then for a symmetry reduction respecting the minimal axioms (\ref{eqn:infinitesimalconditions}), it is necessary that
the following Lie bracket relations hold true
\begin{flalign}
 \com{V_a}{\mathfrak{g}}\subseteq\mathfrak{g}~,\forall_a~.
\end{flalign}
In other words, $(\mathrm{span}(t_i,V_a),\com{~}{~})\subseteq(\Xi,\com{~}{~})$ forms a Lie algebra with ideal $\mathfrak{g}$.
Here $t_i$ are the generators of $\mathfrak{g}$.
\end{propo}
\noindent Note that this gives conditions relating the {\it classical} Lie algebra $(\mathfrak{g},\com{~}{~})$ with the twist.

Next, we evaluate the $\star$-Lie bracket condition (\ref{eqn:infconda}).
Using the explicit form of the inverse twist (\ref{eqn:jsrmatrix}) and (\ref{eqn:condition2}) we obtain
\begin{subequations}
\begin{flalign}
 &\bar f_{\alpha_{(n)}}(t_i^\star) = \com{V_{a_1}}{\cdots\com{V_{a_n}}{t_i^\star}\cdots} = \bigl(\mathcal{N}^\star_{a_n}\cdots\mathcal{N}^\star_{a_1} \bigr)_i^j t_j^\star=:\mathcal{N}^{\star j}_{\alpha_{(n)}i} t_j^\star~,\\
 &\bar f^{\alpha_{(n)}}(t_i^\star) = \Theta^{\beta_{(n)}\alpha_{(n)}} \bar f_{\beta_{(n)}}(t_i^\star)~,\\
 &\Theta^{\beta_{(n)}\alpha_{(n)}} := \frac{1}{n!}\left(\frac{i\lambda}{2}\right)^n \theta^{b_1 a_1} \cdots \theta^{b_n a_n}~,
\end{flalign}
\end{subequations}
where $\alpha_{(n)},\beta_{(n)}$ are multi indices. This leads to
\begin{flalign}
\label{eqn:prestarcommutator}
 \com{t_i^\star}{t_j^\star}_\star = \Theta^{\beta_{(n)}\alpha_{(n)}} \mathcal{N}^{\star k}_{\beta_{(n)} i} \mathcal{N}^{\star l}_{\alpha_{(n)}j}\com{t_k^\star}{t_l^\star}~.
\end{flalign}
Note that in particular for the choice $t_i^\star = t_i, ~\forall_i,$ the $\star$-Lie subalgebra closes with structure constants
\begin{flalign}
\label{eqn:starliealgebra}
 \com{t_i}{t_j}_\star = \Theta^{\beta_{(n)}\alpha_{(n)}} \mathcal{N}^{ k}_{\beta_{(n)} i} \mathcal{N}^{ l}_{\alpha_{(n)}j}\com{t_k}{t_l} = \Theta^{\beta_{(n)}\alpha_{(n)}} \mathcal{N}^{ k}_{\beta_{(n)} i} \mathcal{N}^{ l}_{\alpha_{(n)}j}  f_{kl}^{~~m} t_m =: f_{ij}^{\star~m} t_m~,
\end{flalign}
where we have used the $\mathcal{N}$ defined in (\ref{eqn:condition3}).
This leads to the following
\begin{propo}
\label{propo:starliealgebra}
 Let $\com{V_a}{\mathfrak{g}_\star}\subseteq\mathfrak{g}_\star~,\forall_a$. Then we can always construct a $\star$-Lie subalgebra
 $(\mathfrak{g}_\star,\com{~}{~}_\star)\subseteq(\Xi_\star,\com{~}{~}_\star)$
by choosing the generators as $t_i^\star = t_i$ for all $i$. With this we have $\mathfrak{g}_\star = \mathfrak{g}$ as vector spaces 
and the structure constants are deformed as
\begin{flalign}
 f_{ij}^{\star ~m}=\Theta^{\beta_{(n)}\alpha_{(n)}} \mathcal{N}^{ k}_{\beta_{(n)} i} \mathcal{N}^{ l}_{\alpha_{(n)}j}  f_{kl}^{~~m}~.
\end{flalign}
\end{propo}
\noindent 
Since the condition (\ref{eqn:infcondb}) together with the requirement $t_i^\star=t_i$, for all $i$, automatically fulfills
(\ref{eqn:infconda}), we choose $t_i^\star=t_i$, for all $i$, as a
canonical embedding. In general, other possible embeddings
require further constructions to fulfill condition
(\ref{eqn:infconda}) and are therefore less natural.
We will discuss possible differences between this and other embeddings later on, when we construct the $\star$-Hopf subalgebra and the 
$\star$-Lie derivative action on $\star$-tensor fields.

In addition, we obtain that the necessary condition (\ref{eqn:envelopcond}) for extending $\mathfrak{g}_\star$ to the $\star$-enveloping 
subalgebra $(U\mathfrak{g}_\star,\star)\subseteq(U\Xi_\star,\star)$
 is automatically fulfilled, since  we have $\bar R_{\alpha_{(n)}}(\mathfrak{g}_\star)\subseteq \mathfrak{g}_\star$ 
for all $\alpha_{(n)}$ and additionally
\begin{flalign}
\bar R^{\alpha_{(n)}}(\mathfrak{g}_\star)=(-2)^n \Theta^{\beta_{(n)}\alpha_{(n)}}\bar R_{\beta_{(n)}}(\mathfrak{g}_\star)\subseteq\mathfrak{g}_\star~,~\forall{\alpha}_{(n)}.
\end{flalign}

Next, we evaluate the conditions (\ref{eqn:hopfconditions}), which have to be fulfilled in order to construct the $\star$-Hopf subalgebra 
$\mathcal{H}_\mathfrak{g}^\star\subseteq\mathcal{H}_{\Xi}^\star$. For the particular choice of the twist (\ref{eqn:jstwist})
 we obtain the following
\begin{propo}
\label{propo:starhopf}
Let $(U\mathfrak{g}_\star,\star)\subseteq(U\Xi_\star,\star)$ be a $\star$-enveloping subalgebra and let the deformation parameter
 $\lambda\neq 0$.
Then in order to extend $(U\mathfrak{g}_\star,\star)$ to the $\star$-Hopf subalgebra $\mathcal{H}_\mathfrak{g}^\star=(U\mathfrak{g}_\star,\star,\Delta_\star,S_\star,\epsilon_\star)\subseteq\mathcal{H}_\Xi^\star$ the 
condition
\begin{flalign}
 V_{a_1} \in \mathfrak{g}_\star~,\quad \text{if  }~ \com{V_{a_2}}{\mathfrak{g}_\star}\neq\lbrace0\rbrace
\end{flalign}
has to hold true for all pairs of indices $(a_1,a_2)$ connected by the antisymmetric matrix $\theta$ (\ref{eqn:theta}), i.e.~$(a_1,a_2)\in \big\lbrace(1,2),(2,1),(3,4),(4,3),\dots\big\rbrace$.
\end{propo}
\noindent Note that these conditions depend on the embedding $t_i^\star=t_i^\star(t_j)$. 
The proof of this proposition is shown in the appendix \ref{app:proof}.

Finally, if we demand $\mathcal{H}_\mathfrak{g}^\star$ to be a triangular $\star$-Hopf algebra (\ref{eqn:triangularcond}) we obtain the stringent condition
\begin{flalign}
 V_a \in \mathfrak{g}_\star~,~\forall_a~.
\end{flalign}
This can be shown by using $X_{R_{\alpha}} = R_\alpha$ and $V_a \star V_b = V_a V_b$, which holds true for the class of Reshetikhin-Jambor-Sykora twists.

As we have seen above, there are much stronger restrictions on the Lie algebra $(\mathfrak{g},\com{~}{~})$ 
and the twist, if we want to extend the deformed infinitesimal transformations $(\mathfrak{g}_\star,\com{~}{~}_\star)$ 
to the (triangular) $\star$-Hopf subalgebra $\mathcal{H}_\mathfrak{g}^\star$. In particular this extension restricts the $V_a$ themselves,
 while for infinitesimal transformations and the finite transformations $(U\mathfrak{g}_\star,\star)$ only the images of $V_a$ 
acting on $\mathfrak{g}_\star$ are important.

Next, we study the $\star$-action of the $\star$-Lie and Hopf algebra on the deformed tensor fields. The $\star$-action of the generators
 $t_i^\star$ on $\tau\in\mathcal{T}_\star$ is defined by (\ref{eqn:starliederivative}) and simplifies to
\begin{flalign}
 \mathcal{L}^\star_{t_i^\star}(\tau) = \Theta^{\alpha_{(n)}\beta_{(n)}} \mathcal{N}^{\star j}_{\alpha_{(n)}i} ~\mathcal{L}_{t_j^\star} \bigl(\bar f_{\beta_{(n)}}(\tau)\bigr)~.
\end{flalign}
For invariant tensors, the $\star$-Lie derivative has to vanish to all orders in $\lambda$, since we work with formal power series.
If we now for explicitness take the natural choice $t_i^\star=t_i$ we obtain the following
\begin{propo}
\label{propo:starinvariance}
 Let $\com{V_a}{\mathfrak{g}_\star}\subseteq\mathfrak{g}_\star~,\forall_a$ and $t_i^\star=t_i,~\forall_i$. Then a tensor $\tau\in\mathcal{T}_\star$ is $\star$-invariant
under $(\mathfrak{g}_\star,\com{~}{~}_\star)$, if and only if it is invariant under the undeformed action of $(\mathfrak{g},\com{~}{~})$, i.e.
\begin{flalign}
 \mathcal{L}^{\star}_{\mathfrak{g}_\star}(\tau)=\lbrace 0\rbrace~\Leftrightarrow~\mathcal{L}_{\mathfrak{g}}(\tau)=\lbrace 0\rbrace~.
\end{flalign}
\end{propo}
\begin{proof}
 For the proof we make the ansatz $\tau=\sum\limits_{n=0}^\infty \lambda^n \tau_{n}$ and investigate $\mathcal{L}^\star_{t_i}(\tau)$
order by order in $\lambda$, since we work with formal power series. By using (\ref{eqn:condition3}) to reorder the
Lie derivatives such that $t_i$ is moved to the right, it can be shown recursively in powers of $\lambda$ that the proposition 
holds true.
\end{proof}
\noindent Note that for $t_i^\star\neq t_i$ this does not necessarily hold true. We can not make statements for this case, since
we would require a general solution of (\ref{eqn:prestarcommutator}), which we do not have yet. But we mention again that
we consider choosing $t_i^\star$ different from $t_i$ quite unnatural.

This proposition translates to the case of finite symmetry transformations with $t_i^\star=t_i$ because of the properties of
the $\star$-Lie derivative.

The framework developed in this section will now be applied to
cosmology and black holes in order to give some specific examples and
discuss possible physical implications.

\section{\label{sec:cosmo}Application to Cosmology}
In this section we will investigate models with symmetry group $E_3$ in four spacetime dimensions with topology $\mathbb{R}^4$. 
These are flat Friedmann-Robertson-Walker (FRW) universes.
The undeformed Lie algebra of this group is generated by the ``momenta'' $p_i$ and ``angular momenta'' $L_i$, $i\in\lbrace1,2,3\rbrace$, 
which we can represent in the Lie algebra of vector fields as
\begin{flalign}
 p_i = \partial_i\quad,\quad L_i = \epsilon_{ijk}x^j \partial_k~,
\end{flalign}
where $\epsilon_{ijk}$ is the Levi-Civita symbol.

The undeformed Lie bracket relations are
\begin{flalign}
 \com{p_i}{p_j}=0~,\quad\com{p_i}{L_j}=-\epsilon_{ijk}p_k~,\quad\com{L_i}{L_j}=-\epsilon_{ijk}L_k~.
\end{flalign}

We will work with the natural embedding $t_i^\star=t_i$, and therefore the $\star$-Lie subalgebra is given by
 $\mathfrak{g}_\star=\mathfrak{e}_{3\star}=\mathfrak{e}_3=\mathrm{span}(p_i,L_i)$.

We can now explicitly evaluate the condition each twist vector field $V_a$ has to satisfy given by
 $\com{V_a}{\mathfrak{e}_{3}}\subseteq\mathfrak{e}_{3}$ (cf.~proposition \ref{propo:ideal}). 
Since the generators are at most linear in the spatial coordinates, $V_a$ can be at most quadratic in order to 
fulfill this condition. If we make a quadratic ansatz with time dependent coefficients we obtain that each $V_a$ 
has to be of the form
\begin{flalign}
\label{eqn:FRWV}
 V_a = V_a^0(t) \partial_t + c_a^i\partial_i + d_a^i L_i + f_a x^i\partial_i~,
\end{flalign}
where $c_a^i,~d_a^i,~f_a\in\mathbb{R}$ and $V_a^0(t)\in C^\infty(\mathbb{R})$ in order to obtain hermitian deformations.
If all $V_a$ have the form (\ref{eqn:FRWV}), the $\star$-Lie algebra closes (cf.~proposition \ref{propo:starliealgebra}).

Next, we have to find conditions such that the $V_a$ are mutually commuting. A brief calculation shows that the
following conditions have to be fulfilled:
\begin{subequations}
\label{eqn:frwcond}
\begin{align}
 \label{eqn:frwcond1}&d_a^i d_b^j \epsilon_{ijk} = 0 ~,\forall_k~,\\
 \label{eqn:frwcond2}&c_a^i d_b^j \epsilon_{ijk} - c_b^i d_a^j \epsilon_{ijk} + f_a c_b^k - f_b c_a^k =0~,\forall_k ~,\\
 \label{eqn:frwcond3}&\com{V_a^0(t)\partial_t}{V_b^0(t)\partial_t}=0~.
\end{align}
\end{subequations}
As a first step, we will now work out all possible deformations of $\mathfrak{e}_{3}$ when twisted with two commuting vector fields. 
We will classify the possible solutions. Therefore we divide the solutions into classes depending on the value of $d_a^i$ and $f_a$.
 We use as notation for our cosmologies $\mathfrak{C}_{AB}$, where $A\in\lbrace1,2,3\rbrace$ and $B\in\lbrace1,2\rbrace$, 
which will become clear later on, when we sum up the results in table \ref{tab:frw}.

Type $\mathfrak{C}_{11}$ is defined to be vector fields with $d_1^i=d_2^i=0$ and $f_1=f_2=0$, i.e.~
\begin{flalign}
V_{1(\mathfrak{C}_{11})} = V_{1}^0(t)\partial_t +c_{1}^i\partial_i~,\quad V_{2(\mathfrak{C}_{11})} = V_{2}^0(t)\partial_t +c_{2}^i\partial_i~.
\end{flalign}
These vector fields fulfill the first two conditions (\ref{eqn:frwcond1}) and (\ref{eqn:frwcond2}). The solutions of the third condition 
(\ref{eqn:frwcond3}) will be discussed later, since this classification we perform now does not depend on it.

Type $\mathfrak{C}_{21}$ is defined to be vector fields with $d_1^i=d_2^i=0$, $f_1\neq0$ and $f_2=0$.
 The first condition (\ref{eqn:frwcond1}) is trivially fulfilled and the second (\ref{eqn:frwcond2}) is fulfilled, 
if and only if $c_{2}^i=0,~\forall_i$, i.e.~type $\mathfrak{C}_{21}$ is given by the vector fields
\begin{flalign}
 \tilde V_{1(\mathfrak{C}_{21})} = V_{1}^0(t)\partial_t + c_{1}^i\partial_i + f_{1} x^i\partial_i~,\quad \tilde V_{2(\mathfrak{C}_{21})} = V_{2}^0(t)\partial_t~.
\end{flalign}
These vector fields can be simplified to
\begin{flalign}
 V_{1(\mathfrak{C}_{21})} = c_{1}^i\partial_i + f_{1} x^i\partial_i~,\quad V_{2(\mathfrak{C}_{21})} = V_{2}^0(t)\partial_t~,
\end{flalign}
since both lead to the same twist (\ref{eqn:jstwist}).

Solutions with $d_1^i=d_2^i=0$, $f_1\neq0$ and $f_2\neq0$ lie in type $\mathfrak{C}_{21}$, since we can perform the twist conserving map
$V_2\to V_2 - \frac{f_2}{f_1} V_1$, which transforms $f_2$ to zero. Furthermore $\mathfrak{C}_{31}$ is defined by 
$d_1^i=d_2^i=0$, $f_1=0$ and $f_2\neq0$ and is equivalent to $\mathfrak{C}_{21}$ by interchanging the labels of the vector fields.

Next, we go on to solutions with without loss of generality $\mathbf{d}_1\neq0$ and $\mathbf{d}_2=0$ ($\mathbf{d}$ denotes the vector). 
Note that this class contains also the class with $\mathbf{d}_1\neq0$ and $\mathbf{d}_2\neq0$.
 To see this, we use the first condition (\ref{eqn:frwcond1}) and obtain that $\mathbf{d}_1$ and $\mathbf{d}_2$ have to be parallel, 
i.e.~$d^i_2=\kappa d^i_1$.
Then we can transform $\mathbf{d}_2$ to zero by using the twist conserving map $V_2\to V_2 - \kappa V_1$.

Type $\mathfrak{C}_{12}$ is defined to be vector fields with $\mathbf{d}_1\neq0$, $\mathbf{d}_2=0$ and $f_1=f_2=0$.
The first condition (\ref{eqn:frwcond1}) is trivially fulfilled, while the second condition (\ref{eqn:frwcond2}) 
requires that $\mathbf{c}_2$ is parallel to $\mathbf{d}_1$, i.e.~we obtain
\begin{flalign}
V_{1(\mathfrak{C}_{12})} = V_{1}^0(t)\partial_t + c_{1}^i\partial_i + d_{1}^i L_i~,\quad V_{2(\mathfrak{C}_{12})} = V_{2}^0(t)\partial_t + \kappa~d_{1}^i\partial_i~,
\end{flalign}
where $\kappa\in\mathbb{R}$ is a constant.

Type $\mathfrak{C}_{22}$ is defined to be vector fields with $\mathbf{d}_1\neq0$, $\mathbf{d}_2=0$, $f_1\neq0$ and $f_2=0$.
Solving the second condition (\ref{eqn:frwcond2}) (therefore we have to use that the vectors are real!) we obtain
\begin{flalign}
 V_{1(\mathfrak{C}_{22})} = c_{1}^i\partial_i + d_{1}^i L_i + f_{1} x^i\partial_i~,\quad V_{2(\mathfrak{C}_{22})} = V_{2}^0(t)\partial_t~,
\end{flalign}
where we could set without loss of generality $V_{1}^0(t)$ to zero, as in type $\mathfrak{C}_{21}$. Note that $\mathfrak{C}_{21}$ is
 contained in $\mathfrak{C}_{22}$ by violating the condition $\mathbf{d}_1\neq0$.

Finally, we come to the last class, type $\mathfrak{C}_{32}$, defined by $\mathbf{d}_1\neq0$, $\mathbf{d}_2=0$, $f_1=0$ and $f_2\neq0$. 
This class contains also the case $\mathbf{d}_1\neq0$, $\mathbf{d}_2=0$, $f_1\neq0$ and $f_2\neq0$ by using the twist
 conserving map $V_1\to V_1 - \frac{f_1}{f_2} V_2$. The vector fields are given by
\begin{flalign}
 V_{1(\mathfrak{C}_{32})} = V_{1}^0(t)\partial_t + \frac{d_{1}^j c_{2}^k\epsilon_{jki}}{f_2}  \partial_i + d_{1}^i L_i ~,\quad V_{2(\mathfrak{C}_{32})} = V_{2}^0(t)\partial_t + c_{2}^i\partial_i + f_{2} x^i\partial_i~.
\end{flalign}

Note that type $\mathfrak{C}_{11}$ and $\mathfrak{C}_{12}$ can be extended to a triangular $\star$-Hopf 
subalgebra by choosing $V_1^0(t)=V_2^0(t)=0$ in each case.

For a better overview we additionally present the the results in table \ref{tab:frw}, containing all possible two vector field 
deformations $\mathfrak{C}_{AB}$ of the Lie algebra of the euclidian group. 
From this table the notation $\mathfrak{C}_{AB}$ becomes clear.
\begin{table}
\begin{center}
\begin{tabular}{|l|l|l|}
\hline
{\large $~~\mathfrak{C}_{AB}$} & $\mathbf{d}_1=\mathbf{d}_2=0$ & $\mathbf{d}_1\neq0$~,~$\mathbf{d}_2=0$ \\ \hline
$f_1=0$,    &$V_1= V_{1}^0(t)\partial_t +c_{1}^i\partial_i$  &$V_1= V_{1}^0(t)\partial_t + c_{1}^i\partial_i + d_{1}^i L_i$   \\ 
$f_2=0$	   &$V_2= V_{2}^0(t)\partial_t +c_{2}^i\partial_i$  &$V_2= V_{2}^0(t)\partial_t + \kappa~d_{1}^i\partial_i$	     \\ \hline
$f_1\neq0$, &$V_1= c_{1}^i\partial_i + f_{1} x^i\partial_i$  &$V_1= c_{1}^i\partial_i + d_{1}^i L_i + f_{1} x^i\partial_i$    \\ 
$f_2=0$	   &$V_2= V_{2}^0(t)\partial_t$	                    &$V_2= V_{2}^0(t)\partial_t$	                             \\ \hline
$f_1=0$,    &$V_1= V_{1}^0(t)\partial_t$	                    &$V_1= V_{1}^0(t)\partial_t + \frac{1}{f_2}d_{1}^j c_{2}^k\epsilon_{jki}  \partial_i + d_{1}^i L_i$	\\ 
$f_2\neq0$ &$V_2= c_{2}^i\partial_i + f_{2} x^i\partial_i$  &$V_2= V_{2}^0(t)\partial_t + c_{2}^i\partial_i + f_{2} x^i\partial_i$	\\ \hline
\end{tabular}
\end{center}
\caption{\label{tab:frw}Two vector field deformations of the cosmological symmetry group $E_3$.}
\end{table}

Next, we discuss solutions to the third condition (\ref{eqn:frwcond3}) $\com{V_1^0(t)\partial_t}{V_2^0(t)\partial_t}=0$. 
It is obvious that choosing either 
$V_1^0(t)=0$ or $V_2^0(t)=0$ and the other one arbitrary is a solution. Additionally, we consider solutions with $V_1^0(t)\neq0$ and $V_2^0(t)\neq0$.
 Therefore there has to be some point $t_0\in\mathbb{R}$, such that without loss of generality $V_1^0(t)$ is unequal zero in some open region 
$U\subseteq\mathbb{R}$ around $t_0$. 
In this region we can perform the diffeomorphism $t\to \tilde t(t):= \int\limits_{t_0}^t dt^\prime \frac{1}{V_1^0(t^\prime)}$ 
leading to $\tilde V_1^0(\tilde t)=1$. With this the third condition (\ref{eqn:frwcond3}) becomes
\begin{flalign}
 0=\com{V_1^0(t)\partial_t}{V_2^0(t)\partial_t}=\com{\tilde V_1^0(\tilde t)\partial_{\tilde t}}{\tilde V_2^0(\tilde t)\partial_{\tilde t}}
 = \Bigl(\partial_{\tilde t}\tilde V_2^0(\tilde t)\Bigr) \partial_{\tilde t}~.
\end{flalign}
This condition is solved if and only if $\tilde V_2^0(\tilde t) =\mathrm{const.}$ for $t\in U\subseteq\mathbb{R}$. 
For the subset of analytical functions $C^\omega(\mathbb{R})\subset C^\infty(\mathbb{R})$ we can continue this condition 
to all $\mathbb{R}$ and obtain the global relation $V_2^0(t)= \kappa V_1^0(t)$, with some constant $\kappa\in\mathbb{R}$.
For non analytic, but smooth functions, we can not continue these relations to all $\mathbb{R}$ and therefore only obtain local conditions  restricting the functions in the overlap of their supports to be linearly dependent. 
In particular non analytic functions with disjoint supports fulfill the condition (\ref{eqn:frwcond3}) trivially.

After characterizing the possible two vector field deformations of $\mathfrak{e}_3$ we briefly give a method how to obtain twists
 generated by a larger number of vector fields. For this purpose we use the canonical form of $\theta$ (\ref{eqn:theta}).

Assume that we want to obtain deformations with e.g.~four vector fields. Then of course all vector fields have to be of the form (\ref{eqn:FRWV}). 
According to the form of $\theta$ we have two blocks of vector fields $(a,b)=(1,2)$ and $(a,b)=(3,4)$, in which the classification described
 above for two vector fields can be performed. This means that all four vector field twists can be obtained by using two types
 of two vector field twists.
 We label the twist by using a tuple of types, e.g.~$(\mathfrak{C}_{11},\mathfrak{C}_{22})$ means that $V_1,V_2$ are of type $\mathfrak{C}_{11}$ and $V_3,V_4$ of type $\mathfrak{C}_{22}$.
But this does only assure that $\com{V_a}{V_b}=0$ for $(a,b)\in\lbrace(1,2),(3,4)\rbrace$ and we have to demand further restrictions in
 order to fulfill $\com{V_a}{V_b}=0$ for all $(a,b)$ and that all vector fields give independent contributions 
to the twist. In particular twists constructed with linearly dependent vector fields can be reduced to a twist 
constructed by a lower number of vector fields.

This method naturally extends to a larger number of vector fields, until we cannot find anymore independent and
 mutually commuting vector fields. We will now give two examples for the $\mathfrak{e}_3$ case in order to clarify the construction.

As a first example we construct the four vector field twist $(\mathfrak{C}_{11},\mathfrak{C}_{11})$. 
In this case all four vector fields commute without imprinting further restrictions. We assume that three of the four vectors
 $\mathbf{c}_a$ are linearly independent, such that the fourth one, say $\mathbf{c}_4$, can be decomposed into the other ones.
 If we now choose four linearly independent functions $V_a^0(t)$ (this means that they are non analytic) leads to a proper four vector field
 twist.

As a second simple example we construct the four vector field twist $(\mathfrak{C}_{21},\mathfrak{C}_{21})$. 
In order to have commuting vector fields we obtain the condition $c_3^i=\frac{f_3}{f_1}c_1^i$.
We therefore have $V_3=\frac{f_3}{f_1}V_1$ and the four vector field twist can be reduced to the two vector field twist
 of type $\mathfrak{C}_{21}$ with $\tilde V_1 =V_1$ and $\tilde V_2 = V_2 +\frac{f_3}{f_1}V_4$. This is an example 
of an improper four vector field twist.

This method can be applied in order to investigate general combinations of two vector field twists, if one requires them.
 Because this construction is straightforward and we do not require these twists for our discussions, we do not present them here.

At the end we calculate the $\star$-commutator of the linear coordinate functions $x^\mu\in A_\star$ for the various types of models
 in first order in the deformation parameter $\lambda$. It is given by
\begin{flalign}
 c^{\mu\nu}:=\starcom{x^\mu}{x^\nu} := x^\mu\star x^\nu -x^\nu\star x^\mu = i\lambda \theta^{ab} V_a(x^\mu) V_b(x^\nu) +\mathcal{O}(\lambda^2)~.
\end{flalign}
The results are given in appendix \ref{app:starcom} and show that these commutators can be at most quadratic in 
the spatial coordinates $x^i$.
Possible applications of these models will be discussed in the outlook, see section \ref{sec:conc}.

\section{\label{sec:blackhole}Application to Black Holes}
In this section we investigate possible deformations of non rotating black holes. 
We will do this in analogy to the cosmological models and therefore do not have to explain every single step.

The undeformed Lie algebra of the symmetry group $\mathbb{R}\times SO(3)$ of a non rotating black hole is generated by
 the vector fields
\begin{flalign}
p^0=\partial_t\quad,\quad L_i = \epsilon_{ijk} x^j\partial_k~,
\end{flalign}
given in cartesian coordinates. We choose $t_i^\star =t_i$ for all $i$ 
and define $\mathfrak{g}_\star=\mathfrak{g} = \mathrm{span}(p^0,L_i)$.

It can be shown that each twist vector field $V_a$ has to be of the form
\begin{flalign}
V_a = (c^0_a(r)+N_a^0 t)\partial_t + d_a^i L_i + f_a(r) x^i \partial_i
\end{flalign}
in order to fulfill $\com{V_a}{\mathfrak{g}}\subseteq\mathfrak{g}$. Here $r=\Vert\mathbf{x}\Vert$ is the euclidian 
norm of the spatial position vector.

The next task is to construct the two vector field deformations. Therefore we additionally have to
 demand $\com{V_a}{V_b} =0,~\forall_{a,b}$, leading to the conditions
\begin{subequations}
\begin{flalign}
\label{eqn:blackcond1}&d_a^i d_b^j\epsilon_{ijk}=0~,~\forall_k~, \\
\label{eqn:blackcond2}&(f_a(r) x^j \partial_j - N_a^0) c^0_b(r) - (f_b(r) x^j\partial_j -N_b^0) c^0_a(r)  =0~,\\
\label{eqn:blackcond3}& f_a(r)f_b^\prime(r)-f_a^\prime(r) f_b(r) = 0~,
\end{flalign}
\end{subequations}
where $f_a^\prime(r)$ means the derivative of $f_a(r)$. 
Note that (\ref{eqn:blackcond3}) is a condition similar to (\ref{eqn:frwcond3}), and therefore
 has the same type of solutions. Because of this, the functions $f_1(r)$ and $f_2(r)$ have to be parallel in the overlap 
of their supports. From this we can always eliminate locally one $f_a(r)$
by a twist conserving map and simplify the investigation of the condition (\ref{eqn:blackcond2}). 
At the end, the local solutions have to be glued together.
We choose without loss of generality $f_1(r)=0$ for our classification of local solutions.

The solution to (\ref{eqn:blackcond1}) is that the $\mathbf{d}_a$ have to be parallel. We use
\begin{flalign}
 \mathbf{d}_a=\kappa_a \mathbf{d}
\end{flalign}
with constants $\kappa_a\in\mathbb{R}$ and some arbitrary vector $\mathbf{d}\neq0$.

We now classify the solutions to (\ref{eqn:blackcond2}) according to $N_a^0$ and $f_2(r)$ and label them by $\mathfrak{B}_{AB}$. 
We distinguish between $f_2(r)$ being the zero function or not. The result is shown in table \ref{tab:blackhole}.
Other choices of parameters can be mapped by a twist conserving map into these classes.
Note that in particular for analytical functions $f_a(r)$ the twist conserving map transforming $f_1(r)$ to zero 
can be performed globally, and with this also the classification of twists given in table \ref{tab:blackhole}.
\begin{table}
\begin{center}
\begin{tabular}{|l|l|l|}
\hline
 {\large$~~\mathfrak{B}_{AB}$}  &  $f_2(r)=0$  &  $f_2(r)\neq0$  \\ \hline
$N_1^0=0$,	&$V_1=c_1^0(r)\partial_t +\kappa_1 d^iL_i$	&$V_1=c_1^0\partial_t +\kappa_1 d^i L_i $	\\
$N_2^0=0$	&$V_2=c_2^0(r)\partial_t+\kappa_2 d^i L_i$	&$V_2=c_2^0(r)\partial_t +\kappa_2 d^i L_i + f_2(r) x^i\partial_i$		\\ \hline
$N_1^0\neq0$,	&$V_1=(c_1^0(r)+N_1^0 t)\partial_t $  &$V_1=(c_1^0(r)+N_1^0 t)\partial_t+\kappa_1 d^i L_i$	\\
$N_2^0=0$	&$V_2=\kappa_2 d^i L_i$				&$V_2= -\frac{1}{N_1^0}f_2(r)r c_1^{0\prime}(r)\partial_t +\kappa_2 d^i L_i+f_2(r)x^i\partial_i$\\ \hline
$N_1^0=0$,	&$V_1=\kappa_1 d^i L_i$				&$V_1=c_1^0(r) \partial_t + \kappa_1 d^i L_i$,\quad\text{with (\ref{eqn:blackode})} \\
$N_2^0\neq0$	&$V_2=(c_2^0(r)+N_2^0 t)\partial_t $  &$V_2= (c_2^0(r)+N_2^0 t)\partial_t +\kappa_2 d^i L_i +f_2(r) x^i\partial_i $\\ \hline
\end{tabular}
\end{center}
\caption{\label{tab:blackhole}Two vector field deformations of the black hole symmetry group $\mathbb{R}\times SO(3)$. 
Note that $c_1^0(r)=c_1^0$ has to be constant in type $\mathfrak{B}_{12}$.}
\end{table}

In type $\mathfrak{B}_{32}$ we still have to solve a differential equation for $c_1^0(r)$ given by
\begin{flalign}
\label{eqn:blackode}
 c_1^0(r) = \frac{f_2(r)}{N_2^0} r c_1^{0\prime}(r)~,
\end{flalign}
for an arbitrary given $f_2(r)$.
We will not work out the solutions to this differential equation, since type $\mathfrak{B}_{32}$ is a quite unphysical model,
 in which the noncommutativity is increasing linear in time due to $N_2^0\neq 0$.

Note that $\mathfrak{B}_{11}$ can be extended to a triangular $\star$-Hopf algebra by choosing $c_a^0(r)=c_a^0$, for $a\in\lbrace1,2\rbrace$.
In addition, $\mathfrak{B}_{12}$ is a $\star$-Hopf algebra for $\kappa_1=\kappa_2=0$.

The $\star$-commutators $c^{\mu\nu}=\starcom{x^\mu}{x^\nu}$ of the coordinate functions $x^\mu\in A_\star$ in order $\lambda^1$ for these 
models are given in the appendix \ref{app:starcom}. They can be used in order to construct sensible
 physical models of a noncommutative black hole.

By using the method explained in the previous section, the two vector field twists can be extended to multiple vector field
 twists. Since we do not require these twists in our work and their construction is straightforward, we do not
 present them here.

\section{\label{sec:conc}Conclusion and Outlook}
We have discussed symmetry reduction in noncommutative gravity using
the formalism of twisted noncommutative differential geometry.  Our
motivation for these investigations derives from the fact that, for
most physical applications of gravity theories, including cosmology,
symmetry reduction is required due to the complexity of such models,
already in the undeformed case.

In section~\ref{sec:symred} we have presented a general method for
symmetry reduction in twisted gravity theories.  As a result we have
obtained restrictions on the twist, depending on the structure of the
twisted symmetry group.  In particular, we find that deforming the
infinitesimal symmetry transformations results in weaker restrictions
than deforming the finite transformations and demanding a quantum
group structure.  In section~\ref{sec:jambor} we have applied this
general method to gravity theories twisted by Reshetikhin-Jambor-Sykora twists.
These are twists constructed from commuting vector fields.  In this
case we could give explicit conditions, which have to be fulfilled in
order to allow symmetry reduction of a given Lie group.

In sections~\ref{sec:cosmo} and~\ref{sec:blackhole} we have
investigated admissible deformations of FRW and black hole symmetries
by a Reshetikhin-Jambor-Sykora twist. In this class we have classified all
possible deformations.  This lays the foundation for phenomenological
studies of noncommutative cosmology and black hole physics based on
twisted gravity.

In a forthcoming work~\cite{Ohl/Schenkel:Cosmo:2009} we will
investigate cosmological implications of twisted FRW models by
studying fluctuations of quantum fields living on twisted FRW
backgrounds. Quantum fields were already introduced in a twisted
framework in~\cite{Aschieri:2007sq}. As we see from
proposition~\ref{propo:starinvariance}, the noncommutative backgrounds
are also invariant under the undeformed action of the classical
symmetry. This means that they have the same coordinate
representations with respect to the undeformed basis vectors as the
commutative fields in Einstein gravity. With this we have a construction
principle for noncommutative backgrounds, in their natural basis,
by representing the classical fields in the deformed basis.
A class of models of particular interest is
type~$\mathfrak{C}_{22}$ in section~\ref{sec:cosmo}
(cf.~table~\ref{tab:frw}).  These twists break classical translation
invariance, but classical rotation invariance can be retained by
tuning~$\mathbf{d}_1$ and~$\mathbf{c}_1$ to small values. Furthermore,
the global factor~$V_2^0(t)$ in the exponent of the twist can be used
in order to tune noncommutativity effects depending on time.
Obviously, enforcing a suitable~$V_2^0(t)$ by hand leads to
phenomenologically valid models.

Since there is no natural choice of~$V_2^0(t)$, it is interesting to
investigate the dynamics of~$V_2^0(t)$ in a given field configuration
and study if it leads to a model consistent with cosmological
observations. In this case, the model would be physically
attractive. This will also be subject of future
work~\cite{Ohl/Schenkel:Cosmo:2009}.  Dynamical noncommutativity has
already been studied in the case of scalar field theories on Minkowski
spacetime~\cite{Aschieri:2008zv}.

In the case of black hole physics, models of particular interest would
be~$\mathfrak{B}_{11}$ with functions~$c^0_a(r)$ decreasing
sufficiently quickly with~$r$ and~$\mathfrak{B}_{12}$ with~$f_2(r)$
and~$c_2^0(r)$ decreasing sufficiently quickly with~$r$
(cf.~table~\ref{tab:blackhole}).  It will again be interesting to
investigate the dynamics of these functions on a given field
configuration.  Note that the type~$\mathfrak{B}_{12}$
with~$\kappa_1=\kappa_2=0$ is invariant under the classical black hole
symmetries, and therefore particularly interesting for physical
applications.  On the other hand, models with nonvanishing~$N_a^0$ are
of little physical interest, because the noncommutativity is growing
linearly in time, which would be unphysical.

Other avenues for future work are the classification of models on
nontrivial topologies (like, e.g.,~$\mathbb{R}\times S_3$ in
cosmology), investigating nontrivial
embeddings~$t_i^\star=t_i^\star(t_j)$ and using a wider class of twist
elements.

\section*{Acknowledgements}
AS thanks Christoph Uhlemann and Julian Adamek for discussions and
comments on this work.  This research is supported by Deutsche
Forschungsgemeinschaft through the Research Training Group 1147
\textit{Theoretical Astrophysics and Particle Physics}.

\appendix

\section{\label{app:proof}Proof of Proposition~\ref{propo:starhopf}}
In this appendix we show that for Reshetikhin-Jambor-Sykora twists (\ref{eqn:jstwist}) the conditions (\ref{eqn:hopfconditions}) necessary for
 extending the $\star$-enveloping subalgebra $(U\mathfrak{g}_\star,\star)$ to a $\star$-Hopf subalgebra are
 equivalent to the simplified conditions of proposition \ref{propo:starhopf}. 
The plan is as follows: we use (\ref{eqn:hopfconditions1}) and show that
 it is equivalent to the conditions of proposition \ref{propo:starhopf}. In a second step, we show that (\ref{eqn:hopfconditions2}) 
is automatically satisfied if (\ref{eqn:hopfconditions1}) is fulfilled, and thus does not lead to additional conditions.

We start with (\ref{eqn:hopfconditions1}) and show the identity
\begin{flalign}
 \Delta_\star (U\mathfrak{g}_\star)\subseteq U\mathfrak{g}_\star\otimes U\mathfrak{g}_\star ~\Leftrightarrow~\Delta_\star (\mathfrak{g}_\star)\subseteq U\mathfrak{g}_\star\otimes U\mathfrak{g}_\star~.
\end{flalign}
The direction $\Rightarrow$ is trivial, since $\mathfrak{g}_\star\subset U\mathfrak{g}_\star$, and the direction $\Leftarrow$ can be shown 
 using that $\Delta_\star$ is a $\star$-algebra homomorphism and that $U\mathfrak{g}_\star$ closes under $\star$-multiplication.

Furthermore, using (\ref{eqn:defstarhopfactions}) we obtain
\begin{flalign}
 \Delta_\star (\mathfrak{g}_\star)\subseteq U\mathfrak{g}_\star\otimes U\mathfrak{g}_\star~\Leftrightarrow~X_{\bar R^\alpha}\in U\mathfrak{g}_\star~,~~\text{for all }\alpha\text{ with }\bar R_\alpha(\mathfrak{g}_\star)\neq\lbrace0\rbrace~.
\end{flalign}
Therefore we have to use that all $\bar R^\alpha$ are linearly independent, that $X$ is a vector space isomorphism~\cite{Aschieri:2005zs} 
and that we have $\bar R_\alpha(\mathfrak{g}_\star)\subseteq \mathfrak{g}_\star$, 
due to the minimal axioms (\ref{eqn:infinitesimalconditions}).
 
Additionally, we can show that $X_{\bar R^\alpha}=\bar f^\beta \bar R^\alpha \chi S^{-1}(\bar f_\beta)= \bar R^\alpha$. This is done
 by applying the explicit form of the twist (\ref{eqn:jstwist}) and using that the $V_a$ mutually commute.

Next, we show that 
\begin{flalign}
\label{eqn:prepropo}
 \bar R^\alpha\in U\mathfrak{g}_\star~,~~\text{for all }\alpha\text{ with }\bar R_\alpha(\mathfrak{g}_\star)\neq\lbrace0\rbrace~
\Leftrightarrow~ \theta^{ba}V_b\in\mathfrak{g}_\star~,~~ \text{for all }a\text{ with }\com{V_a}{\mathfrak{g}_\star}\neq\lbrace0\rbrace~.
\end{flalign}
The direction $\Rightarrow$ is trivial, since the RHS is a special case of the LHS. 
The direction $\Leftarrow$ can be shown by using that the $V_a$ mutually commute and the explicit expression of the $R$-matrix 
(\ref{eqn:jsrmatrix}). 

Finally, the RHS of (\ref{eqn:prepropo}) is equivalent to the condition of proposition \ref{propo:starhopf} by 
using the canonical form of $\theta$ (\ref{eqn:theta}).

Next, we show that (\ref{eqn:hopfconditions2}) is satisfied, if (\ref{eqn:hopfconditions1}) is fulfilled. 
For this we use that for Reshetikhin-Jambor-Sykora twists we have $\chi=f^\alpha S(f_\alpha)=1$, which leads to the identity
\begin{flalign}
 S_\mathcal{F}(\xi) = \chi S(\xi)\chi^{-1} =S(\xi) = S^{-1}(\xi) = S_\mathcal{F}^{-1}(\xi)~,~\forall \xi\in U\Xi~ 
\end{flalign}
for the antipode in the $\mathcal{F}$-Hopf algebra.
This property translates to the $\star$-Hopf algebra, since it is isomorphic to the $\mathcal{F}$-Hopf algebra and we obtain
 the following equivalences of (\ref{eqn:hopfconditions2})
\begin{flalign}
\label{eqn:antipodesimple}
 S_\star(U\mathfrak{g}_\star)\subseteq U\mathfrak{g}_\star ~\Leftrightarrow~S_\star(\mathfrak{g}_\star)\subseteq U\mathfrak{g}_\star
~\Leftrightarrow~S^{-1}_\star(\mathfrak{g}_\star)\subseteq U\mathfrak{g}_\star~.
\end{flalign}
For the first equivalence we had to use that $S_\star$ is a $\star$-anti homomorphism.

Using the RHS of (\ref{eqn:prepropo}), which is equivalent to (\ref{eqn:hopfconditions1}), and the definition of $S_\star^{-1}$ 
(\ref{eqn:defstarhopfactions}), we obtain
\begin{flalign}
 S^{-1}_\star(\mathfrak{g}_\star) 
= -\sum_{n=0}^{\infty}\frac{(-i\lambda)^n}{n!} \theta^{a_1b_1}\cdots
\theta^{a_n b_n} \com{V_{a_1}}{\cdots,\com{V_{a_n}}{\mathfrak{g}_\star}\cdots}~ V_{b_1}\cdots V_{b_n} \in U\mathfrak{g}_\star~,
\end{flalign}
where we have used $\xi \star V_a = \xi V_b$ for all $\xi\in U\Xi_\star$, since the action of the twist on $V_a$ is trivial.

\section{\label{app:starcom}$\star$-Commutators of the Coordinate
  Functions in FRW and Black Hole Models} 
In tables~\ref{tab:cosmocom} and~\ref{tab:blackcom}, we list the $\star$-commutators among the linear
coordinate functions to order~$\lambda^1$ in the FRW and black hole
models.  In these expressions, $(i\leftrightarrow j)$ denotes the
same term with~$i$ and~$j$ interchanged.
\begin{table}
\begin{center}
\begin{tabular}{|l|l|}
\hline
Type	&	$c^{\mu\nu}:=\starcom{x^\mu}{x^\nu}$ in $\mathcal{O}(\lambda^1)$\\
\hline
$\mathfrak{C}_{11}$    &	$c^{0i}=i\lambda \bigl( V_{1}^0(t) c_{2}^i - V_{2}^0(t) c_{1}^i  \bigr)$\\
~		&	$c^{ij}=i\lambda \bigl( c_{1}^i c_{2}^j - (i\leftrightarrow j) \bigr)$\\ \hline
$\mathfrak{C}_{21}$    &	$c^{0i}=-i\lambda V_{2}^0(t) \bigl( c_{1}^i + f_{1} x^i \bigr)$\\
~		&	$c^{ij}=0$  \\ \hline
$\mathfrak{C}_{12}$    &	$c^{0i}=i\lambda \bigl( V_{1}^0(t) \kappa d_{1}^i - V_{2}(t) (c_{1}^i+d_{1}^k\epsilon_{kli}x^l )   \bigr)$\\
~		&	$c^{ij}=i\lambda \kappa \bigl( (c_{1}^i +d_{1}^k \epsilon_{kli}x^l ) d_{1}^j - (i\leftrightarrow j)  \bigr)$  \\ \hline
$\mathfrak{C}_{22}$    &	$c^{0i}=-i\lambda V_{2}^0(t) \bigl( c_{1}^i + d_{1}^j\epsilon_{jki}x^k + f_{1} x^i \bigr)$\\
~		&	$c^{ij}=0$  \\ \hline
$\mathfrak{C}_{32}$    &	$c^{0i}=i\lambda \bigl( V_1^0(t) (c_2^i + f_2 x^i) - V_2^0(t) (\frac{1}{f_2}d_1^j c_2^k\epsilon_{jki} + d_1^j\epsilon_{jki}x^k)   \bigr)$\\
~		&	$c^{ij}=i\lambda \bigl( (\frac{1}{f_2} d_1^k c_2^l \epsilon_{kli} +d_1^k\epsilon_{kli} x^l )~(c_2^j + f_2 x^j) - (i\leftrightarrow j)  \bigr)$  \\ \hline
\end{tabular}
\end{center}
\caption{\label{tab:cosmocom} $\star$-commutators in the cosmological models $\mathfrak{C}_{AB}$.}
\end{table}
\begin{table}
\begin{center}
\begin{tabular}{|l|l|}
\hline
Type	&	$c^{\mu\nu}:=\starcom{x^\mu}{x^\nu}$ in $\mathcal{O}(\lambda^1)$\\
\hline
$\mathfrak{B}_{11}$ & $c^{0i}=i\lambda \bigl(c_1^0(r) \kappa_2 -c_2^0(r) \kappa_1\bigr) d^j\epsilon_{jki}x^k$ \\
~		& $c^{ij}=0$ \\ \hline
$\mathfrak{B}_{21}$ & $c^{0i}=i\lambda \bigl(c_1^0(r)+N_1^0 t\bigr) d^j \epsilon_{jki}x^k$\\
~		 & $c^{ij}=0$\\ \hline
$\mathfrak{B}_{12}$ & $c^{0i} =i\lambda \bigl( c_1^0 (\kappa_2 d^j\epsilon_{jki} x^k+f_2(r) x^i) -\kappa_1 c_2^0(r) d^j\epsilon_{jki}x^k\bigr)$  \\
~		& $c^{ij}= i\lambda\bigl(\kappa_1 d^k\epsilon_{kli}x^l (\kappa_2 d^m\epsilon_{mnj} x^n +f_2(r)x^j) -(i\leftrightarrow j)\bigr)$ \\ \hline
$\mathfrak{B}_{22}$ &  $c^{0i}=i\lambda \Bigl((c_1^0(r)+N_1^0 t)(\kappa_2 d^j\epsilon_{jki} x^k +f_2(r)x^i) + \frac{1}{N_1^0} f_2(r)rc_1^{0\prime}(r)\kappa_1d^j\epsilon_{jki}x^k \Bigr)$ \\
~		&  $c^{ij}= i\lambda\bigl(\kappa_1 d^k\epsilon_{kli}x^l (\kappa_2 d^m\epsilon_{mnj} x^n +f_2(r)x^j) -(i\leftrightarrow j)\bigr)$\\ \hline
$\mathfrak{B}_{32}$ &  $c^{0i}=i\lambda \bigl( c_1^0(r)~(\kappa_2 d^j\epsilon_{jki} x^k+f_2(r) x^i) -(c_2^0(r) +N_2^0 t) \kappa_1 d^j\epsilon_{jki}x^k \bigr)$\\
~		&  $c^{ij}=i\lambda\bigl(\kappa_1 d^k\epsilon_{kli}x^l (\kappa_2 d^m\epsilon_{mnj} x^n +f_2(r)x^j) -(i\leftrightarrow j)\bigr)$\\ \hline
\end{tabular}
\end{center}
\caption{\label{tab:blackcom}$\star$-commutators in the black hole models $\mathfrak{B}_{AB}$.}
\end{table}


\end{document}